\documentclass{elsart}
\usepackage[ansinew]{inputenc}
\usepackage[tbtags]{amsmath}
\usepackage{amssymb,eurosym}
\usepackage{graphicx}
\usepackage{natbib}
\newcommand{\MPIPKS}{Max Planck Institute for the Physics of Complex Systems,
N\"othnitzer Str.\ 38, 01187 Dresden, Germany}

\begin{document}
\begin{frontmatter}

\title{Reactions to extreme events: moving threshold model}

\author{Eduardo G.\ Altmann,}\footnote{edugalt@pks.mpg.de}
\author{Sarah Hallerberg,}
\author{Holger Kantz}
\address{\MPIPKS}

\begin{abstract}
In spite of precautions to avoid the harmful effects of extreme events, we
experience recurrently phenomena that overcome
the preventive barriers. These barriers usually increase drastically right after
the occurrence of such extreme events, but steadily decay in their absence. In
this paper we consider a simple model that mimics the evolution of the protection
barriers to study the efficiency of the system's reaction to extreme events and
how it changes our perception of the sequence of extreme events itself. We
obtain that the usual method of fighting extreme events introduces a
periodicity in their occurrence and is generally less efficient than the use
of a constant barrier. On the other hand, it shows a good adaptation
to the presence of slow non-stationarities.
\end{abstract}

\end{frontmatter}

\maketitle

\section{Introduction}\label{sec.introduction}

One important motivation for the unified study of {\em extreme events}
is the concentration of the destructive power of different systems in some 
few rare events.   
Remarkable examples are earthquakes,  extreme weather conditions,
epileptic seizures, heart attacks, stock markets crashes, etc...~\cite{book}.
Extreme events are usually generated by complex dynamics that involves the
coupling between many different dimensions and scales~\cite{kantz}. However, the 
characterization of extreme events is usually done in a single
scientifically or socially relevant observable, like the magnitude of
earthquakes, number of days of drought, or the highest wind speed in 
storms. From the point of view of dynamical systems, which is assumed throughout
this paper, we say that these observables are obtained applying an observation
function to the full phase space of the system.  

In this paper we assume a general perspective for the influence of feedback reactions to
extreme events. These reactions may be consequence of human activities or of
natural feedback loops present in the system~\cite{stauffer}. For concreteness we motivate the problem and we associate our
mathematical model with the case of human reactions to floods in rivers~\cite{pinter}, which is a representative example
of the class of extreme events which we are interested in. 
The land use of the surrounding area of a river is
usually determined through a long period of (non-scientific) observations.
The most natural observable is the maximum height of the water in
a period (e.g., annual maxima), even though observables like the
drainage area or the discharge volume are also commonly used in the scientific
literature. 
After a long period of normality, i.e., when the river does not
overcome the standard preventive barriers, protection is usually neglected, and the barriers
inevitably assume a lower value. On the opposite, after the occurrence of a flood (extreme 
event) a lot of attention and efforts are directed to avoid similar
catastrophes in the future and the barriers thus increase. This is the most
natural unplanned human reaction
to extreme events and will constitute the main motivation for the simplified
model analyzed in this paper. More subtle reactions may affect the observable used to characterize
the system. For instance, excavations in the river and the construction of new
buildings and levees in the floodplains, which also depend on the occurrence
of recent floods, modify the available area of the
river. In this case, the measure of the water level is directly affected by
these human activities and not only by the amount of water in the rivers basin.
An even more drastic human activity can change the dynamics of the
system in the phase space: the construction of a water reservoir upstream can
directly control the level of the waters, or, more indirectly, the
precipitation in a region is influenced by the presence of strong human
activity. 

In summary, the human activities act in a kind of feedback loop with the
occurrence of extreme events and can in principle influence three different
levels on the measurement chain:

\begin{itemize}

\item [(I1)] The preventive barriers, e.g., by increasing the protections
  around the river.

\item [(I2)] The observable used to characterize the system, e.g., by digging the river. 

\item [(I3)] The dynamics in the phase space in a more fundamental way,
  e.g., by constructing water reservoirs.

\end{itemize}

The reactions to extreme events may be planned or involuntary and, correspondingly, the two
fundamental questions are:

\begin{itemize}

\item [(Q1)] Which is the best method in order to reduce the number of extreme events?  

\item [(Q2)] Which is the influence of the feedback reactions on our
  perception and on the occurrence of extreme events? 
\end{itemize}

This paper address questions (Q1) and (Q2) through the analysis of a simple stochastic model that
simulates the feedback reactions (I1) and (I2). It is organized as follows. In
Sec.~\ref{sec.model} we present the model.  The time between successive
extreme events is studied in Sec.~\ref{sec.interval} and the efficiency of
the model in Sec.~\ref{sec.efficiency}. In Sec.~\ref{sec.nonstat}
we discuss how our model adapts to the effect of non-stationarities. Finally,
we summarize our conclusions in Sec.~\ref{sec.conclusion}. 


\section{Moving threshold model}\label{sec.model}


We consider here a simplified model for the feedback reactions to extreme
events that takes into account the main features
discussed in the previous section. A random sequence of events~$\xi_n$ is
taken as a stochastic input to our model and represent the
complex phenomenon measured in the physically relevant observable. We say that an extreme event occurs at
time $n$ when~$\xi_n$ overcomes the value of the
barrier $q_n$, i.e., $\xi_n > q_n$. In this case we expect the new value of the
barrier $q_{n+1}$ to be increased proportionally to the extreme value 
$\xi_n$. On the other hand, if no extreme event occurs in time~$n$, the barriers decrease
to a fraction of its previous value. This decay of the barriers occurs
typically due to the short memory underlying the human activities (forgetting), but it
can also appear naturally, e.g., the decay of immunity after
vaccination~\cite{bio1}, or the increasing vulnerability of forest to wind
gusts due to the growth of trees. The change of the barrier size~$q_{n+1}$
can be thus summarized as 
\begin{equation}\label{eq.model}
q_{n+1} = \left\{\begin{array}{ll}
\mbox{max}\{\alpha \xi_n, \beta q_n\} & \mbox{if } \xi_n > q_n \;, \\
             \beta q_n & \mbox{if } \xi_n \leq q_n \;,\\
	     
\end{array}
\right.
\end{equation}
where formally $0<\beta<1$ and $\alpha>0$. The max in the first
equation can assume the value~$\beta q_n$ only for~$\alpha<\beta$, and is introduced in the
model to avoid the artificial reduction of the barrier after an extreme
event. We study the temporal sequence of extreme 
events as a function of the control parameters~$\alpha,\beta$, with special
interest for the cases~$\beta \lessapprox 1$ and $\alpha \gtrapprox 1$, which
means that an event of the size of the last extreme should not overcome the
threshold in the (near) future.
In principle the dynamics defined by Eq.~(\ref{eq.model}) can be applied to any
time series $\{\xi_n\}$ that does not contain the influence of human
activities. 
In order to avoid further complications of our model we consider
initially~$\{\xi_n\}$ to be a Gaussian delta-correlated random variable  with
$\langle \xi_n \rangle=0$, $\sigma_{\xi}=1$, and thus
$\rho(\xi)=\frac{1}{\sqrt{2\pi}} 
e^{-\frac{\xi^2}{2}}$ and $\langle \xi_n \xi_m \rangle = \delta_{n,m}$, where
$\langle . \rangle$ denotes temporal average.  

It is interesting to compare this model to other simple stochastic models used to
simulate, e.g., the occurrence of earthquakes~\cite{shimazaki}, the spikes in
neurons~\cite{joern}, and paradigmatic examples of stochastic resonance~\cite{stocres}. The novel aspect
of the model studied in this paper is the existence of a threshold that varies
deterministically in time depending only on the previous extreme
events.

It is sometimes convenient to study the dynamics of Eq.~(\ref{eq.model}) using the
variable 
\begin{equation}\label{eq.y}
y_n \equiv \xi_n - q_n \;,
\end{equation}
where extreme events occur for $y_n>0$. The mean value and variance of~$y_n$
can be written as  

\begin{equation}\label{eq.sigmay}
\begin{array}{ll}
\langle y \rangle &= \langle \xi \rangle - \langle q \rangle = -\langle q
\rangle,\\\\
\sigma_y&=\sqrt{\langle y^2
  -\langle y\rangle^2\rangle}=\sqrt{\sigma_{\xi}^2+\sigma_q^2-2\langle\xi q
  \rangle} \\
&= \sqrt{1+\sigma_q^2} > 1,
\end{array}
\end{equation}
by noting that the term $\langle \xi q \rangle$ is zero due to the lack of
correlation between~$\xi_n$ and~$q_n$.

We would like at this point to associate explicitely our model with the perspective of 
floods in rivers mentioned before. Considering the original
variables~$(\xi,q)$, we regard the human
influence restricted to the delimitation of the 
river domain. In this case~$\xi$ could be the water level and~$q$ the size of
the preventive barrier, measured as the maximum acceptable height of the water before  
causing  damage. On the other hand, if we perform the change of
variable~(\ref{eq.y}), we interpret~$y$ as the
departure of the water level from this threshold ($y<0$ below and~$y>0$ above
threshold), while~$\xi$ as the water in the river basin and~$q$ as a measure
of the modification of the river shape due to human activity. We see thus that
the dynamics defined by Eq.~(\ref{eq.model}) models simultaneously reactions~(I1) and~(I2) 
mentioned in the introduction.   

\begin{figure}[!ht]
\centerline{\includegraphics[width=\columnwidth]{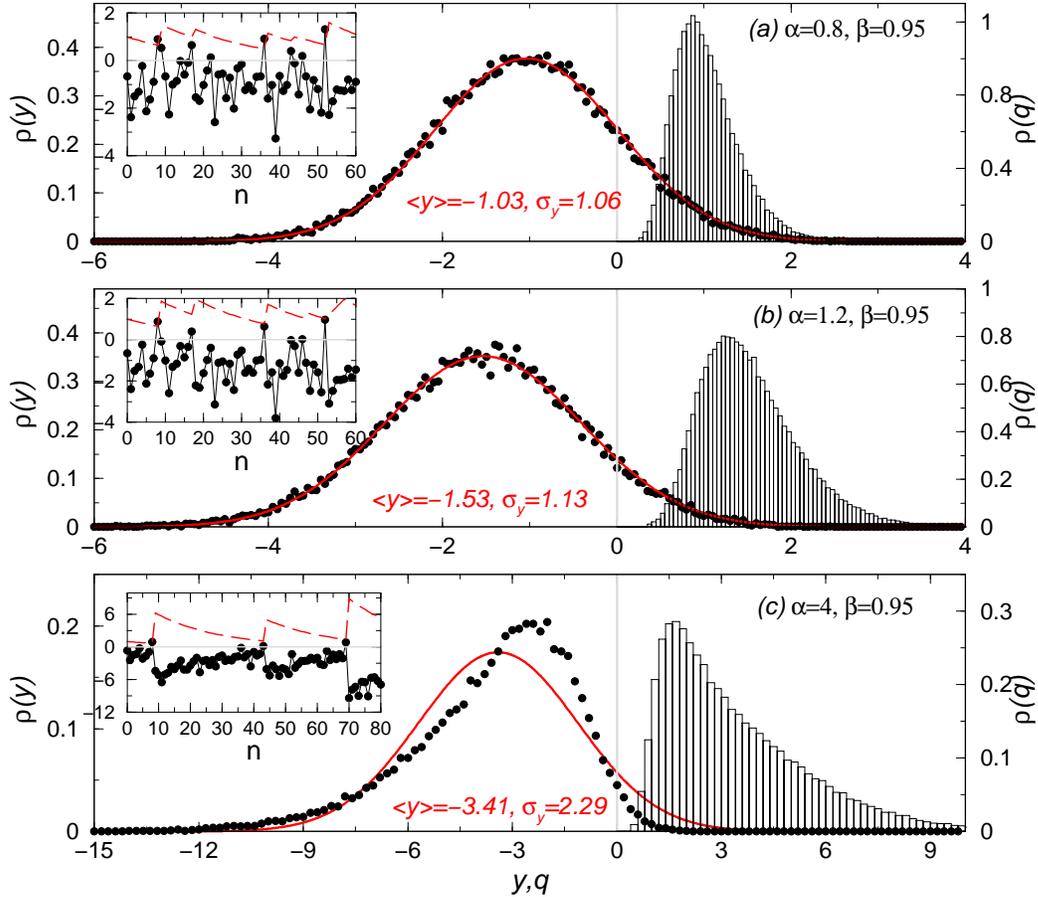}}
\caption{(Color online.) Analysis of the time series of~$y_n=\xi_n-q_n$ and~$q_n$ for three
  different control parameters. For all graphs: black circles represent the PDF
  of~$y$, solid line the Gaussian distribution (with $\langle y
  \rangle=-\langle q \rangle, \sigma=\sigma_y$), and histogram the PDF
  of~$q_n$. In the insets, time series of $q_n$ (dashed line) and $y_n$
  (solid line with circles) are shown. }
\label{fig.series}
\end{figure}

A general picture of our model is presented in 
Fig.~\ref{fig.series}, where numerical results of the time
series~$\{y_n\}$ and~$\{q_n\}$ are shown for three different control parameters~$\alpha$
and~$\beta$. For typical values $(\alpha,\beta\approx1)$, the
probability density function (PDF)~$\rho(q)$ can be
approximated by a Gaussian, what leads to a Gaussian form of
$\rho(y)$. In this case the knowledge of~$\langle y \rangle, \sigma_y$ (see
also Eq.~(\ref{eq.sigmay})) uniquely determines the fraction of extreme events
$\rho(y>0)$. Increasing~$\alpha$ we notice an increase
of$~\sigma_q$ and~$\sigma_y$. For large~$\alpha$, the distribution~$\rho(q)$ becomes highly
asymmetric and a long tail for large $q's$ appears. In this case $\rho(y)$
also loses its normal form. 

\section{Interval between extreme events}\label{sec.interval}

One of the most important characteristics of the temporal sequence of extreme
events is their recurrence time~\cite{recurrence}, i.e., the time~$T$ between 
two successive extreme events quantified by the interevent time distribution~$P(T)$. 
Using a constant barrier~$q_n=q$ the extreme events obtained from an uncorrelated
random time series occur also completely at random and $P(T)$ decays
exponentially. We show in this section that this is not the case when the
size of the barrier dynamically changes according to Eq.(\ref{eq.model}). In
this case~$P(T)$ presents typically a maximum, i.e., there is a characteristic interevent time~$T_{max}>0$.

\begin{figure}[!ht]
\centerline{\includegraphics[width=\columnwidth]{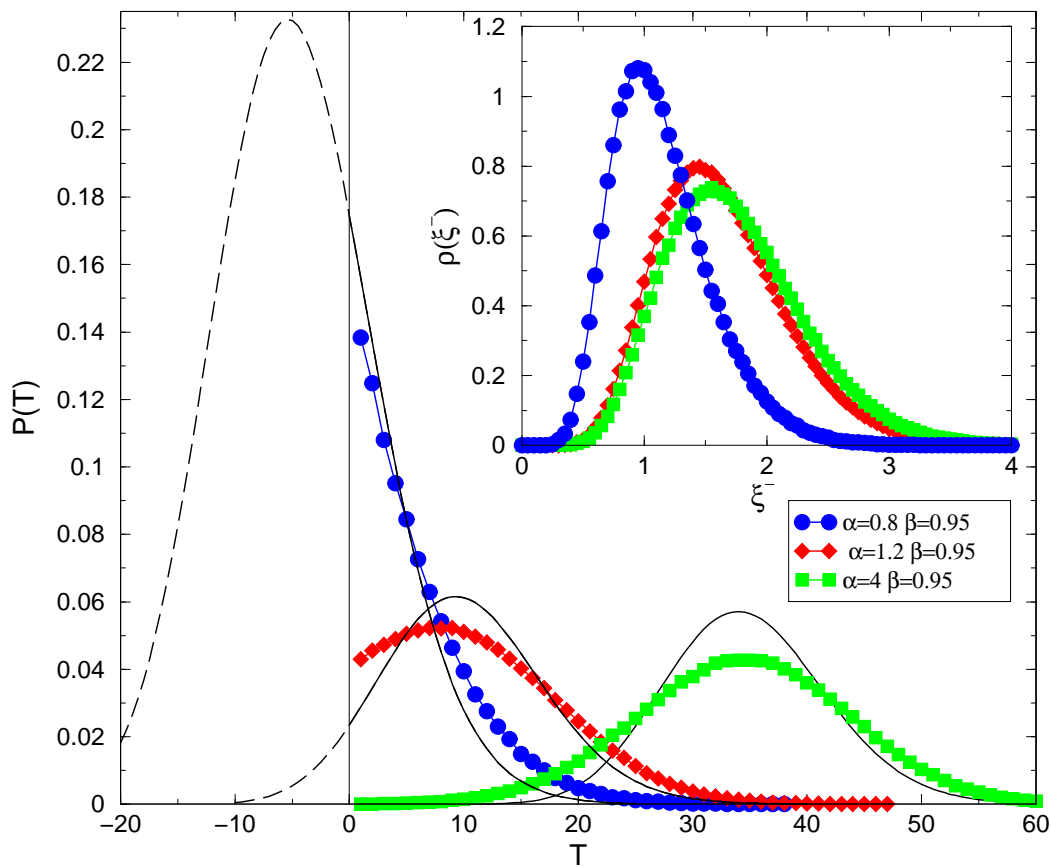}}
\caption{(Color online.) Interevent time distribution $P(T)$ for~$\beta=0.95$ and~$\alpha=0.8,
  1.2,$ and $4$. Lines with symbols are the numerical 
  results and simple lines are the theoretical distribution~(\ref{eq.theoretical})
  using $\xi^-=\langle\xi\rangle=1.17, 1.66, 1.765,$ respectively. After the
  maximum at~$T_{max}$, $P(T)$ decays faster than exponentially to zero (only times~$T>0$ have physical meaning). The inset shows the PDF of~$\xi^-$
  obtained numerically for the same parameters.}
\label{fig.teorico}
\end{figure}

To compute $P(T)$, i.e., the probability of having two {\em consecutive} extreme events
separated by time~$T$, we first have to calculate the probability~$r(t)$ of
having one extreme event at time~$t$ independent of the other events.
In our model the probability of occurrence of one extreme event~$\xi_t>q_t$ at time $t$ is
given by  
\begin{equation}\label{eq.r}
r(t)=\int_{q_t}^\infty \rho(\xi) d\xi=\int_{q_t}^\infty \frac{1}{\sqrt{2 \pi}}
  e^{-\frac{\xi^2}{2}} d\xi =\frac{1}{2}\mbox{erfc}(\frac{\sqrt{2}}{2} q_t), 
\end{equation}
where erfc$(x)$ is the complementary error function. During the interval
between extreme events the barrier evolves as  $q_t=\alpha \xi^- \beta^t$. The
value~$\xi^-$ determines~$q_0=\alpha \xi^-$ and 
corresponds to the value of~$\xi_n$ that generated the
previous extreme event, or,  in the exceptional case  when~$\xi_n > q_n$ but $\beta q_n > \alpha
\xi_n$ [see Eq.~(\ref{eq.model})], to $\xi^-=\beta
q_n/\alpha$. Due to the lack of further correlations between extreme events,
the interevent time distribution~$P(T)$ is obtained as the 
composition of the probability of 
having an extreme event at time~$T$ with the probability that no event occurred for~$t=[0,T[$,
    which can be written as  
$$ P(T) = r(T) \prod_{s=1}^{T-1} [1-r(s)]\approx r(T) \exp[\sum_{s=1}^{T-1} r(s)],$$
where we have used the approximation of small extreme event probability~$r(T) \ll 1$.
In the limit of continuous time we obtain~\cite{stoeckmann}
\begin{equation}\label{eq.pt}
P(T) =C r(T) e^{-\int_0^T r(s) ds},
\end{equation}
where $C$ is a normalization constant. Introducing the expression~(\ref{eq.r})
in~(\ref{eq.pt}) we obtain 
%
{\footnotesize
\begin{equation}\label{eq.theoretical}
P(T;\xi^-)=\frac{C}{2} \mbox{erfc}(\frac{\sqrt{2}}{2} \alpha \xi^-\beta^T)
\exp[-\frac{T}{2}-\frac{\sqrt{2} \alpha \xi^- \beta^T}{2 \ln(\beta) \sqrt{\pi}}
 \;_2F_2( \frac{1}{2},\frac{1}{2};\frac{3}{2},\frac{3}{2};-\frac{1}{2}
(\alpha\xi^-)^2 \beta^{2T})],
\end{equation}
}
%
where $\;_2F_2()$ is the hypergeometric function with
parameters~$a_1=a_2=\frac{1}{2}$
and~$b_1=b_2=\frac{3}{2}$~\cite{hypergeometric}. The interevent time
distribution for given parameters~$\alpha,\beta,$ is given by $P(T)=\int_{0}^{\infty}
P(T;\xi^-) \rho(\xi^-) d\xi^- $. However, the distribution $\rho(\xi^-)$ is
unknown. To obtain simplified theoretical curves we have inserted in
Eq.~(\ref{eq.theoretical}) the constant value $\xi^-=\langle
\xi^-\rangle$, obtained numerically. These distributions are plotted in Fig.~\ref{fig.teorico} where we
notice the existence of a nontrivial most probable interevent time~$T_{max}$,
in good agreement with the numerical results. This constitutes the main and at first sight
most striking result, i.e., extreme events occur almost periodically
if~$\alpha>1$. 

\begin{figure}[!ht]
\centerline{\includegraphics[width=\columnwidth]{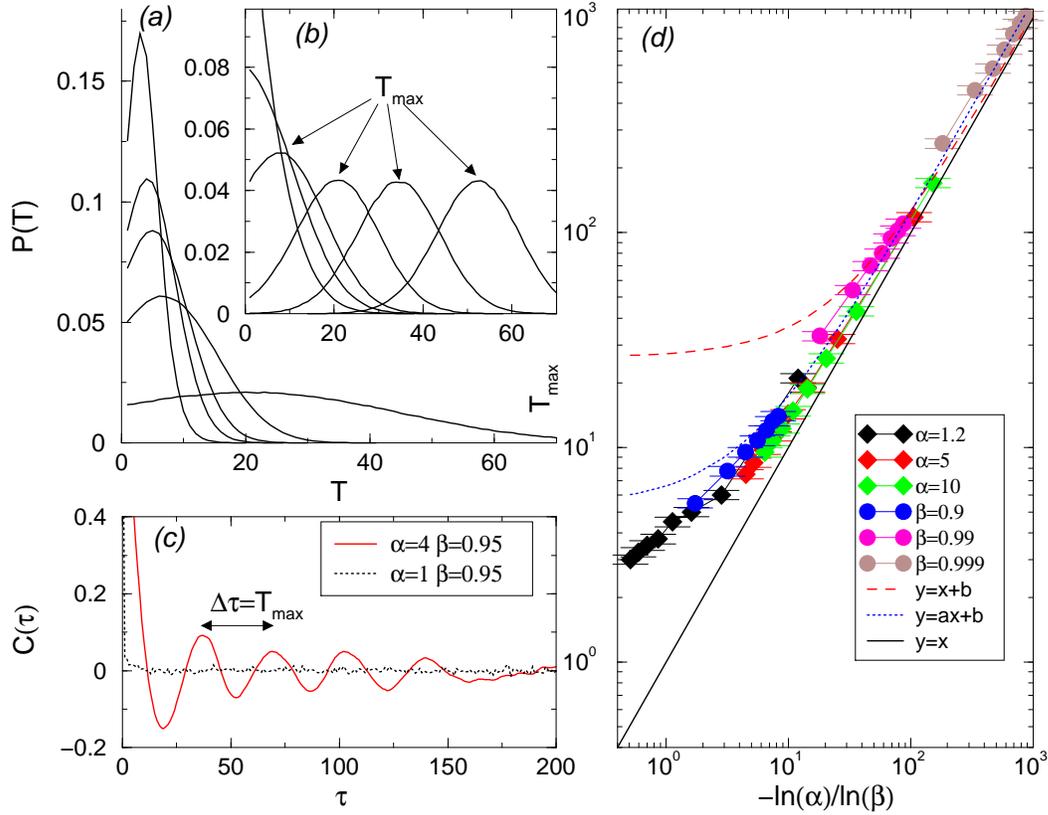}}
\caption{ (Color online.) Numerically obtained interevent time distribution $P(T)$ for (a) $\alpha=1.2$ and
  $\beta=0.7,0.85,0.89,0.94,0.98$ (from left to right), and (b) for $\beta=0.95$ and 
  $\alpha=0.8, 1, 1.2, 2, 4, 10$ (from left to right). (c) Auto-correlation
  function of the series~$y_n$. (d) Dependence of~$T_{max}$ on the
  control parameters~$\alpha,\beta$.}
\label{fig.tempo}
\end{figure}

The existence of such a most probable interevent interval~$T_{max}$ resembles
results 
obtained in models presenting {\em stochastic
  resonance}~\cite{stocres} or {\em coherence
  resonance}~\cite{pikovsky}. However, this is not the case of our model since
it has
neither a periodic input signal nor a resonance behavior for different noise
amplitudes. In fact, in our case~$T_{max}$ varies with the control
parameters~$\alpha,\beta$, whereas a modified (constant in time) variance of ${\xi_n}$ is
equivalent to a simple rescaling of the length scale and do not affect time
scales. We
would like to obtain now the dependence of $T_{max}$ on~$\alpha,\beta$. A 
direct analysis through the theoretical 
distribution~(\ref{eq.theoretical}) is difficult due to its complicated
expression and due to the lack of knowledge of the
distribution of~$\xi^-$, which also strongly depends
on~$\alpha,\beta$. Fortunately some intuition can be gained through
simple approaches developed in what follows. 
Qualitatively, we notice that the existence of a most probable interevent time
is a consequence of the reduction of the probability of short interevent times
due to the increment of the barrier size after one extreme event.
We expect thus
that~$T_{max}$ increases with~$\alpha$ and that for~$\alpha\leq1$, $P(T)$
decays monotonically with~$T$. These results are verified numerically in
Fig.~\ref{fig.tempo}a,b. It is also interesting to note that the characteristic interevent time~$T_{max}$ also
shows up in  the spectrum and autocorrelation function of the series~$\{y_n\}$
and~$\{q_n\}$.  

Consider now that $\xi_n$ assumes the constant
value~$\xi^*>0$. In this simple case the time between
events is given by the time the barrier takes to decay to a value smaller than~$\xi^*$: $q^*=\xi^*=\alpha \xi^* \beta^T\Rightarrow T= -\frac{\ln
  \alpha}{\ln \beta}$. 
Another simple approach is to consider that the next extreme event occurs when the barrier is at~$q^+$
and the previous one occurred due to a value~$\xi^-$, where the distributions
of~($\xi^-,q^+$) are unknown and depend on~$\alpha,\beta$. We obtain in this
case an interevent time~$T$, 
$$q^+=\alpha \xi^-\beta^T \Rightarrow T= -\frac{\ln \alpha}{\ln
  \beta}+\frac{\ln(q^+)-\ln(\xi^-)}{\ln(\beta)}\;.$$
On average $\xi^- > q^+$ and thus $T_{max} > -
\frac{\ln(\alpha)}{\ln(\beta)}$. In both cases we see that 
$T_{max}$ depends explicitely on the ratio
$- \frac{\ln(\alpha)}{\ln(\beta)}$. In Fig.~\ref{fig.tempo}d we plot numerical
obtained values of $T_{max}$ against~$- \frac{\ln(\alpha)}{\ln(\beta)}$. We notice
that all points collapse approximately in a same curve that is indeed always
above the diagonal and that good agreement is achieved by a
linear fitting.

\section{Efficiency of the model}\label{sec.efficiency}

The most relevant issue in the socio-economic context is how to minimize the
number of extreme events, or how to optimize the efficiency of a protection
strategy. 
The rate of extreme events $\rho(y)>0$ is the simplest and most natural
measure of the costs due to extreme events and is thus used throughout this
paper. For many situations it is a realistic measurement of the damage
that occur due to the overpass of a threshold  (e.g,  obstruction of a street or of the power
supply), where there is no difference between "small" and "large" extreme
events. In other situations, more detailed accounts of the  costs would make use of cost functions
that depend (non-linearly) on $y$. 
Our model can be considered as a specific method which can be optimized by the
choice of the control parameters $\alpha,\beta$.

\begin{figure}[!ht]
\centerline{\includegraphics[width=\columnwidth]{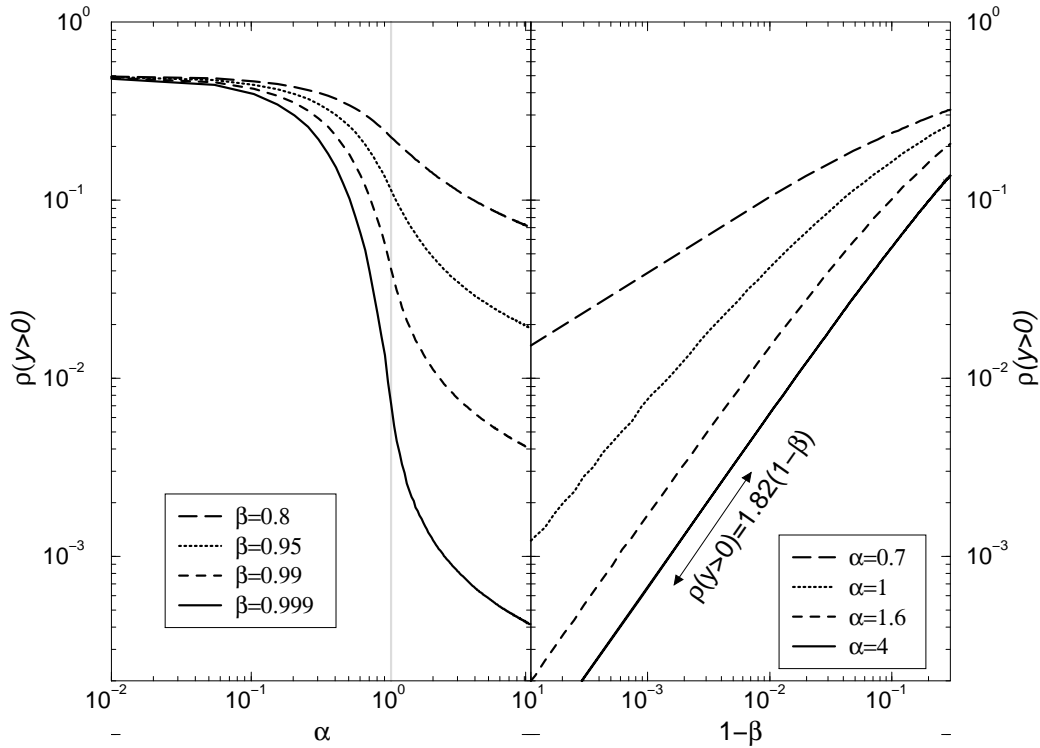}}
\caption{ Extreme events rate~$\rho(y>0)$ as a function of the control
  parameters $\alpha$ and $\beta$. }
\label{fig.parameters}
\end{figure}

The rate of extreme
events~$\rho(y>0)$ as a function of the control parameters is shown in 
Fig.~\ref{fig.parameters}. Since the barrier~$q$ do not necessarily increase after one
extreme event if~$\alpha<1$, we notice again a qualitatively different 
behavior for~$\alpha<1$ and~$\alpha>1$. For a fixed~$\beta$ and
varying~$\alpha$ (Fig.~\ref{fig.parameters}a) we notice that the number of
extreme events decays drastically around~$\alpha\approx1$. On the other hand, by
fixing~$\alpha$ and varying~$\beta$ (Fig.~\ref{fig.parameters}b) the number of
extreme events goes much faster to
zero when~$\beta \rightarrow 1$ if~$\alpha>1$.

\begin{figure}[!ht]
\centerline{\includegraphics[width=\columnwidth]{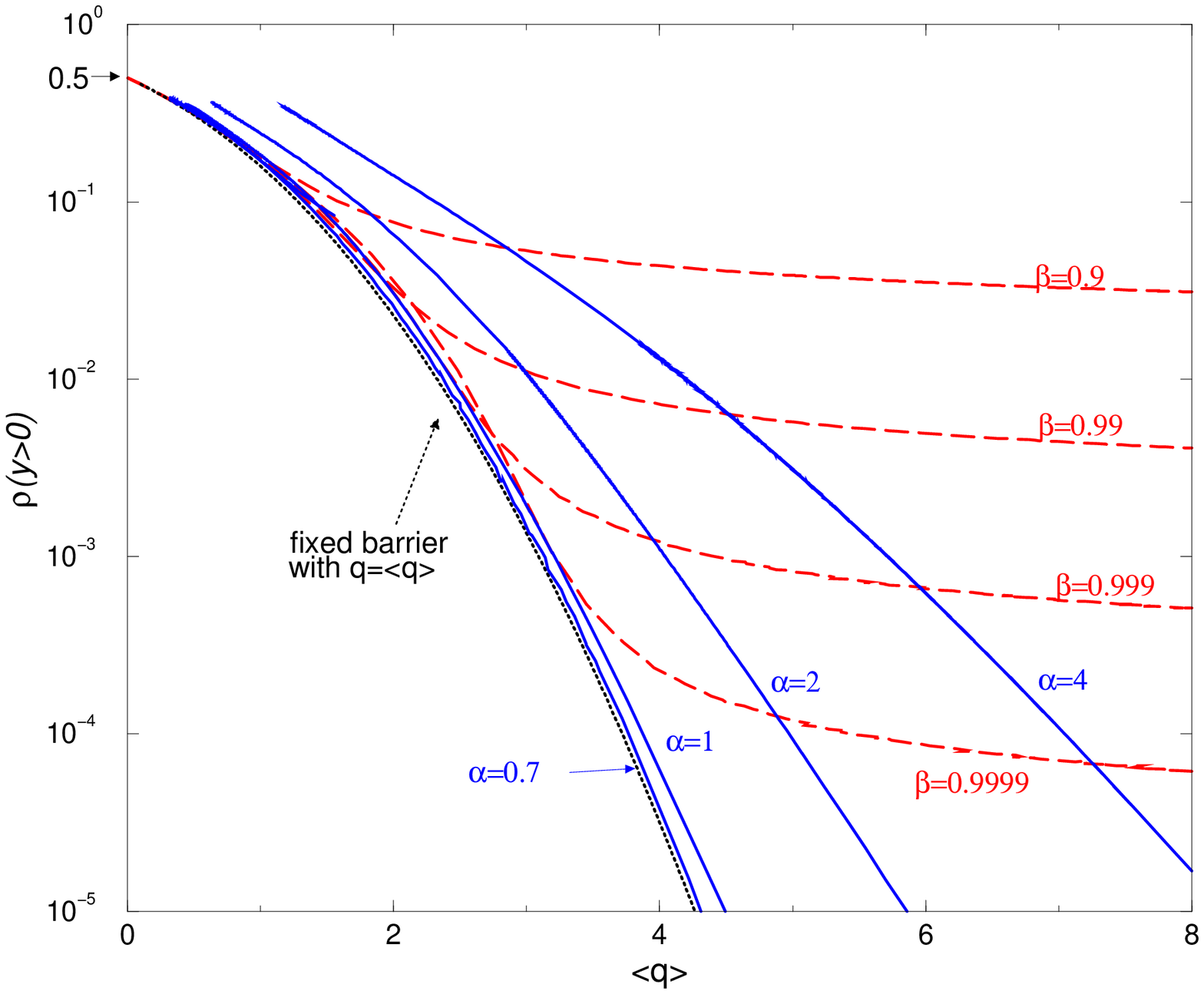}}
\caption{(Color online.) Extreme events rate~$\rho(y>0)$ as a function of the mean
  barrier~$\langle q \rangle$ (costs). Dashed lines are obtained by
  fixing the control parameter~$\beta$ and varying~$\alpha$, and continuous
  lines by fixing~$\alpha$ and varying $\beta$. The dotted line at left
  correspond to a constant barrier in time~$q_t=\langle q \rangle$. }
\label{fig.efficiency}
\end{figure}

It is quite natural that the number of extreme events is reduced
when the control parameters increase. However, in real
situations the increment of these parameters, or equivalently the increment of
the barrier, is related to some costs that have to be taken into account when
studying the efficiency of the model. Since the costs are usually increasing
with the size of the barrier, we measure them by the mean value of the
barriers~$\langle q_n \rangle$.
In Fig.~\ref{fig.efficiency} the rate of extreme events is shown against~$\langle q \rangle$ for different values of the control
parameters~$\alpha,\beta$ and is compared with the result (dotted line)
obtained when the barrier
is maintained unchanged in time. 
As already suggested in Eq.~(\ref{eq.sigmay}), we notice that our moving
threshold method is always less efficient than the constant barrier case. The
limit of unchanged barrier is obtained in our model 
for~$\alpha\rightarrow 0$ and~$\beta \rightarrow 1$, when the most
efficient results are obtained. By noting that the lines of constant~$\alpha$ are
approximately parallel to this limit, we  realize that the relevant limit
is~$\beta\rightarrow1$. Indeed, for any~$\alpha\approx1$ an efficient
reduction of the number of extreme events is only possible by
increasing~$\beta$ towards one.

\section{Non-stationarities}\label{sec.nonstat}

In the previous section we have seen that the model proposed in this paper to
simulate the feedback reactions to extreme events is always less efficient than maintaining the size of the barrier
constant in time, i.e., a non-reactive model. On the other hand, a clear
advantage of reactive models is their ability to deal with non-stationarities
in the time series. This is a specially important issue when considering
extreme events since in many cases they are indeed originated from process
presenting slow trends. Once more, these trends may be natural or consequence of
human activities (e.g., change in land use, global warming).

\begin{figure}[!ht]
\centerline{\includegraphics[width=\columnwidth]{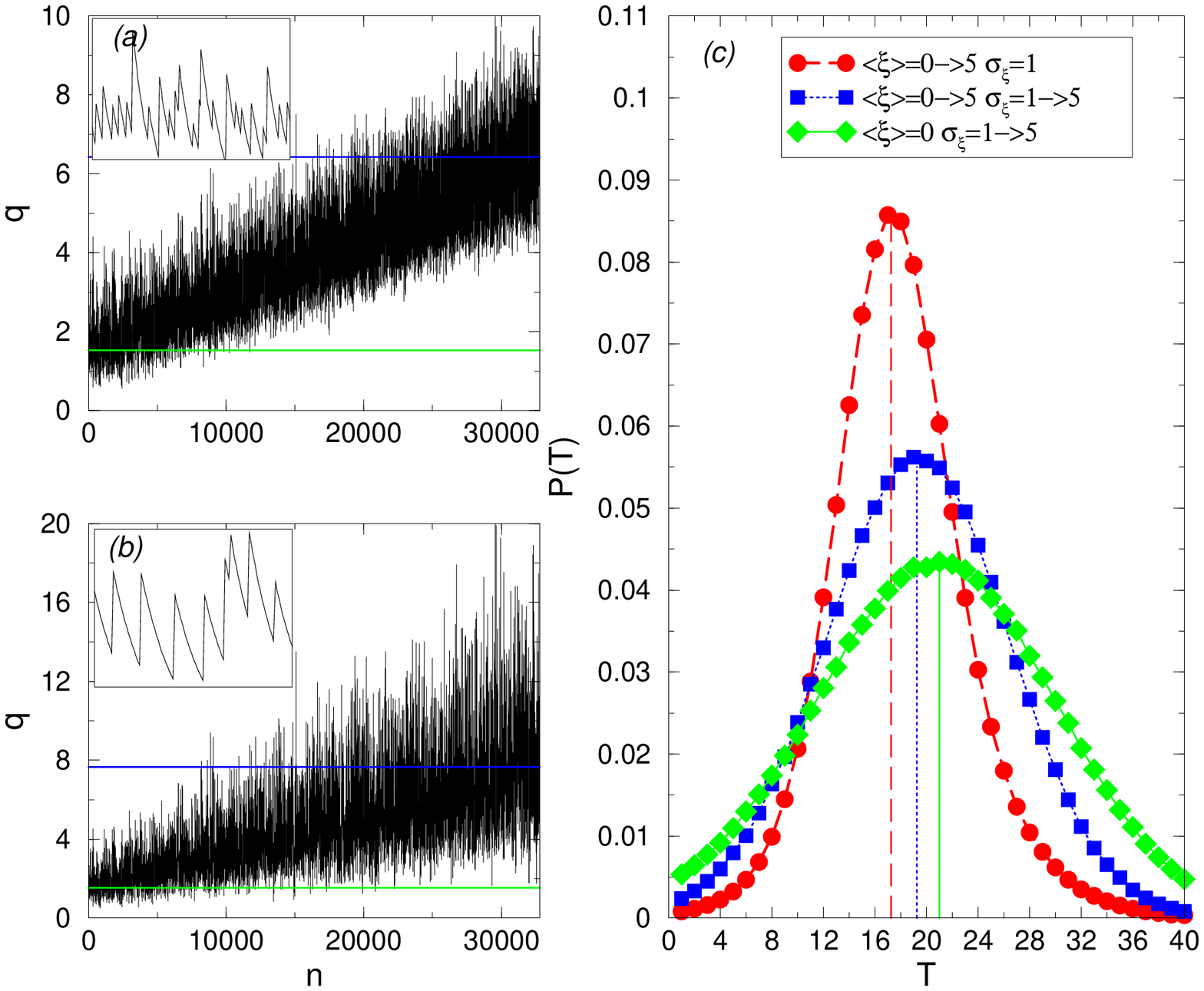}}
\caption{(Color online.) Time series of the barrier $q$
  for~$\alpha=1.2,\beta=0.95$ and: (a) linear increment in time of
  $\langle \xi \rangle$ with $\langle \xi
\rangle^f=5$ and (b) linear increment in time
  of~$\sigma_\xi$ with $\sigma^f_\xi=5$. The horizontal lines correspond
  to~$\langle q \rangle$ at $n=0$ 
  (bottom) and at $n=N=2^{15}$ (top). The insets show magnifications. (c) Numerically obtained interevent 
  time distributions for~$\alpha=2,\beta=0.95$ and different
  non-stationary scenarios (see legend). Vertical lines are located at the value of~$T_{max}$. The stationary case for~$\alpha=2,\beta=0.95$
  is indistinguishable from the case where only $\sigma_\xi$ increases.}
\label{fig.nonstat}
\end{figure}

In order to explore how the model defined by Eq.~(\ref{eq.model}) adapts to
weak non-stationarities, we choose in this section the input time 
series~$\{\xi_n\}$ to be a Gaussian delta correlated random  variable with mean
and variance changing linearly in time
\begin{equation}\label{eq.nonstat}
\begin{array}{ll}
\langle \xi \rangle =& \frac{\langle \xi \rangle^f}{N} n, \\
\sigma_\xi  =& 1+\frac{\sigma^f_\xi-1}{N} n,\\
\end{array}
\end{equation}
where $N$ is the total observation time and $\langle \xi
\rangle^f,\sigma^f_\xi>0$ are constants. Note that increasing the value of $\langle \xi
\rangle$ is {\em not} equivalent to a simple translation since the barrier is still limited to positive values~$q>0$ and
$\lim_{n \rightarrow \infty} \beta q^n=0$.
In Fig.~\ref{fig.nonstat}a,b we show that on large time scales the value of the barrier~$q_n$
also increases linearly in time in both cases, i.e., when the mean or the
variance increases in time.  The linear increment
of $\langle q \rangle$ increase also the fluctuations ($\sigma_q$) of the
time series~$\{q_n\}$ and~$\{y_n\}$.

It is also interesting to compare the results for the interevent time
distribution of non-stationary time series with those reported in 
Sec.~\ref{sec.interval} for the stationary case. When $\langle \xi_n \rangle$ increases (decreases) we
note that the value~$T_{max}$ of the peak of the interevent time
distribution~$P(T)$ slightly decreases (increases).  This effect is quite
natural since the probability of a large value of~$\xi$ is constantly
increasing (decreasing) when $\langle \xi \rangle$ increases (decreases). On the other
hand, due to the linearity of Eqs.~(\ref{eq.model}), a change of the
variance~$\sigma_\xi$ lead to a rescale of~$q$ without changing the value
of~$T_{max}$. Both effects are verified numerically in Fig.~(\ref{fig.nonstat})c.

\begin{figure}[!ht]
\centerline{\includegraphics[width=\columnwidth]{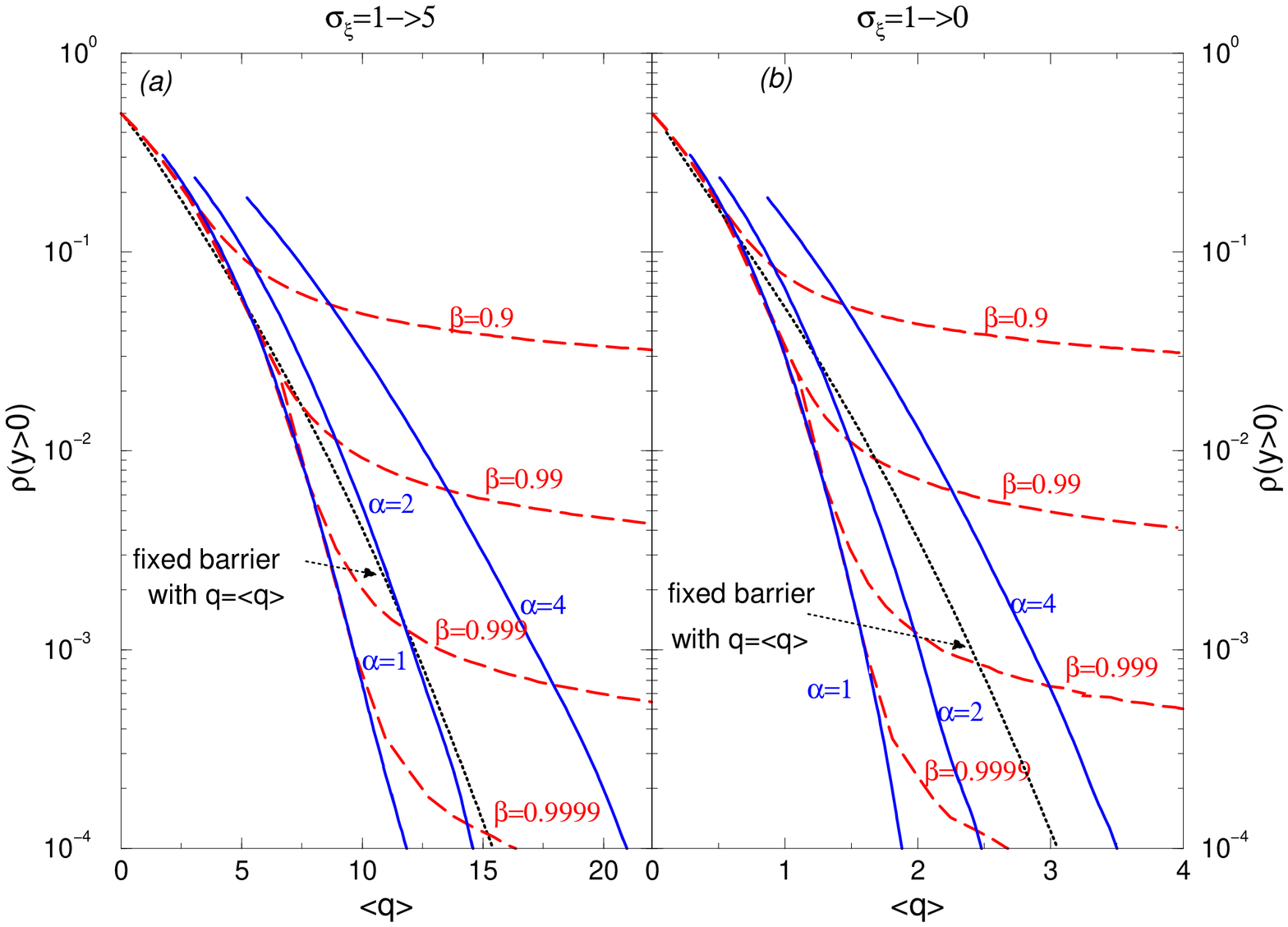}}
\caption{(Color online.) Extreme events rate~$\rho(y>0)$ as a function of the mean
  barrier~$\langle q \rangle$. Dashed lines are obtained by
  fixing the control parameter~$\beta$ and varying~$\alpha$, and continuous
  lines by fixing~$\alpha$ and varying $\beta$. The dotted line 
  correspond to a constant barrier in time~$q_t=\langle q \rangle$. Two
  non-stationary situation are considered (a) $\sigma^f_\xi=5$ and (b)
  $\sigma^f_\xi=0$. $\langle \xi \rangle=0$ is constant for both cases.}
\label{fig.nonstat2}
\end{figure}

Since the barrier increases proportionally to the size of the extreme event, we see that
the time our model takes to adjust to non-stationarities is given by the
interevent time~$T$. When~$T_{max}$ is much smaller than the total observation
time, as considered here, the
non-stationary effects can be considered small during this time interval. 
As a consequence of this fast adaptability of our
model to the application of relations~(\ref{eq.nonstat}) we have that the
dependence of the number of extreme events on the parameters~$\alpha,\beta$ is
qualitatively equivalent to the one reported in Sec.~\ref{sec.efficiency} for
the stationary case. In order to obtain the efficiency we have to compare
again the
extreme events rate~$\rho(y>0)$ with the mean barrier~$\langle q \rangle$
(costs). However, now the value of~$q$ is driven by  changes of~$\langle
\xi \rangle$ (the value~$\langle q \rangle$ reflects only the period
of larger~$\langle \xi \rangle$) showing that the efficiency analysis does not
make sense in this case.  More
interesting is the case when the variance~$\sigma_\xi$ changes in time and the
mean value~$\langle \xi \rangle$ is kept constant, shown in Fig.~\ref{fig.nonstat2} for both
increasing and decreasing~$\sigma_\xi$. The comparison with the results
obtained with constant barrier shows that with reasonable choice of 
parameters~$\alpha,\beta$ the moving threshold model leads to a much more
efficient result.

\section{Conclusion}\label{sec.conclusion}
In summary, we have introduced a model to simulate human reactions or 
natural feedback response to extreme events. We have obtained that the
sequence of extreme events occur 
with a certain periodicity, exclusively due to the human activity (natural
feedback). Regarding 
the efficiency of the model, we have obtained that the best strategy in order
to efficiently reduce the number of extreme events is to try to avoid the decrease of the protection barriers in the periods
between extreme events. On the other hand, if slow non-stationarities are present
in the phenomena, it is also useful to increase the usual protections to the value
of the previous extreme event. The same conclusion is also expected for positively
correlated sequences of events.

These results are obtained in a
very simplified model that tries to isolate the influence of the human reactions to extreme events. In more realistic setups the properties
discussed here may appear together with system-specific characteristics. In
this sense, we can relate the characteristic interevent time observed in our
model with the observed interepidemic interval between, e.g., smallpox
epidemics~\cite{bio2}. Direct association of realistic preventive schemes with the control parameters of our
model lead to the estimation of the characteristic interevent time. For
instance, in the example of floods in river we may estimate that
barriers are reduced by~$2\%$ every year and that after a flood they are
increased by~$20\%$ more than the highest water level. With these parameters
and ignoring deviations of Gaussianity, we obtain through our model the
reasonable estimation of a~$53$ years period between floods.

\section*{Acknowledgments} 
 We thank E. Ullner for helpful discussions.
E.G.A. was supported by CAPES (Brazil) and DAAD (Germany).



\begin{thebibliography}{10}


\bibitem{book}
S. Albeverio, V. Jentsch, H. Kantz (eds.),
\newblock Extreme Events in Nature and Society,
\newblock Springer, Berlin, 2005.


\bibitem{kantz}
H. Kantz {\em et al.},
\newblock Dynamical Interpretation of Extreme events: predictability and
predictions,
\newblock Chapter 4 of Ref. \cite{book}. 

\bibitem{stauffer}
\newblock The opposite effect, i.e., the influence of extreme events in human
opinions, was considered in: 
\newblock Fortunato and Stauffer, Chapter 11 of Ref. \cite{book}.

\bibitem{pinter}
N. Pinter,
\newblock Science 308 (2005) 207.

\bibitem{bio1}
K. Glass and B.T. Grenfell,
\newblock J. theor. Biol. 221 (2003) 121.

\bibitem{shimazaki}
K. Shimazaki and T. Nakata,
\newblock Geophys. Res. Lett. 7 (1980), 279. 



\bibitem{joern}
J. Davidsen and H. G. Schuster,
\newblock Phys. Rev. E 65 (2002), 026120.

\bibitem{stocres}
L. Gammaitoni, P. Hänggi, P. Jung, F. Marchesoni,
\newblock Rev. Mod. Phys. 70 (1998) 223.


\bibitem{recurrence}
E. G. Altmann and H. Kantz,
\newblock Phys. Rev. E 71 (2005) 056106.


\bibitem{stoeckmann} 
H.-J. Stöckmann,
\newblock Quantum Chaos: An introduction,
\newblock Cambridge University Press, Cambridge, 1999 (p. 93).

\bibitem{hypergeometric}
The following identity was used (see http://functions.wolfram.com/):
{\footnotesize
\begin{equation*}
\;_2F_2(a_1,a_2;b_1,b_2;z)=\prod_{k=1}^2 (\frac{\Gamma(b_k)}{\Gamma(a_k)\Gamma(b_k-a_k)})
\int_0^1 \int_0^1 \prod_{k=1}^2 t_k^{a_k-1} (1-t_k)^{-a_k+b_k-1} e^{z t_1
  t_2} dt_1 dt_2. 
\end{equation*}}


\bibitem{pikovsky}
A. S. Pikovsky and J. Kurths,
\newblock Phys. Rev. Lett., 78 (1997) 775.

\bibitem{bio2}
C.J. Duncan, S.R. Duncan, S. Scott,
\newblock J. theor. Biol. 183 (1996) 447.


\end{thebibliography}
\end{document}